\documentclass[preprint, amsmath, amssymb, preprintnumbers, superscriptaddress, floatfix, showpacs, showkeys, aps, prl]{revtex4-1}

\usepackage{graphicx}
\usepackage{float}
\usepackage{color}

\begin{document}

\title{Dirac Cone Protected by Non-Symmorphic Symmetry and 3D Dirac Line Node in ZrSiS}

\author{Leslie M. Schoop}
\email{l.schoop@fkf.mpg.de}
\affiliation{Max Planck Institute for Solid State Research, Heisenbergstr. 1, 70569 Stuttgart, Germany}
\author{Mazhar N. Ali}
\affiliation{Max Plank Institute for Microstructure Physics, Weinberg 2, 06120 Halle, Germany}
\affiliation{IBM-Almaden Research Center, 650 Harry Road, San Jose, CA 95120, USA}
\author{Carola Stra\ss er}
\affiliation{Max Planck Institute for Solid State Research, Heisenbergstr. 1, 70569 Stuttgart, Germany}
\author{Viola Duppel}
\affiliation{Max Planck Institute for Solid State Research, Heisenbergstr. 1, 70569 Stuttgart, Germany}
\author{Stuart S. P. Parkin}
\affiliation{Max Plank Institute for Microstructure Physics, Weinberg 2, 06120 Halle, Germany}
\affiliation{IBM-Almaden Research Center, 650 Harry Road, San Jose, CA 95120, USA}
\author{Bettina V. Lotsch}
\email{b.lotsch@fkf.mpg.de}
\affiliation{Max Planck Institute for Solid State Research, Heisenbergstr. 1, 70569 Stuttgart, Germany}
\affiliation{Department of Chemistry, Ludwig-Maximilians-Universit\"at M\"unchen, Butenandtstr. 5-13, 81377 M\"unchen, Germany}
\affiliation{Nanosystems Initiative Munich (NIM) \& Center for Nanoscience, Schellingstr. 4, 80799 M\"unchen, Germany}
\author{Christian R. Ast}
\affiliation{Max Planck Institute for Solid State Research, Heisenbergstr. 1, 70569 Stuttgart, Germany}
\date{\today}


\begin{abstract}
\textbf{Materials harboring exotic quasiparticles, such as Dirac and Weyl fermions\cite{xu2015discovery,borisenko2015time,weng2015weyl,xu2015observation}, have garnered much attention from the physics and material science communities. These fermions are massless and, in some materials, have shown exceptional physical properties such as ultrahigh mobility and extremely large magnetoresistances \cite{liang2015ultrahigh,ali2014large,du2015unsaturated,shekhar2015large}. Recently, new materials have been predicted to exist which exhibit line nodes of Dirac cones \cite{PhysRevLett.115.036806,xie2015new,burkov2011topological,rhim2015landau}. Here, we show with angle resolved photoemission studies supported by \textit{ab initio} calculations that the highly stable, non-toxic and earth-abundant material, ZrSiS, has an electronic band structure that hosts several Dirac cones which form a Fermi surface with a diamond-shaped line of Dirac nodes. We also experimentally show, for the first time, that the square Si lattice in ZrSiS is an excellent template for realizing the new types of 2D Dirac cones recently predicted by Young and Kane \cite{young2015dirac} and image an unforseen surface state that arises close to the 2D Dirac cone. Finally, we find that the energy range of the linearly dispersed bands is as high as 2\,eV above and below the Fermi level; much larger than of any known Dirac material so far. This makes ZrSiS a very promising candidate to study the exotic behavior of Dirac electrons, or Weyl fermions if a magnetic field is applied, as well as the properties of lines of Dirac nodes.}
\end{abstract}


\maketitle

\indent{} The electronic structure of a three-dimensional (3D) Dirac semi-metal (DSM) contains two sets of linear, doubly degenerate bands which cross at a four-fold degenerate crossing called a Dirac point, a sort of 3D analogue of graphene \cite{ali2014crystal,young2012dirac}. If inversion symmetry (IS) or time reversal symmetry (TRS) are broken, those doubly degenerate bands become spin split, resulting in singly degenerate band crossings called Weyl nodes \cite{wan2011topological,vafek2014dirac}. Although many different materials have been predicted to host Dirac or Weyl fermions \cite{weng2015weyl,gibson2015three,young2012dirac}, only a few real materials have been experimentally verified. Both Cd$_3$As$_2$ and Na$_3$Bi have symmetry protected 3D Dirac cones, which have been imaged with angle resolved photoelectron spectroscopy (ARPES)\cite{neupane2014observation,PhysRevLett.113.027603,liu2014discovery,xu2015observation}. Both materials exhibit exotic transport properties such as ultrahigh mobility, large, linear magnetoresistance and negative magnetoresistance at low fields \cite{xiong2015anomalous,liang2015ultrahigh}. Signatures of a chiral anomaly in ZrTe$_5$ have been seen in ARPES as well as transport experiments \cite{li2014observation}. Weyl fermions have been shown to exist in the IS-breaking compounds TaAs \cite{xu2015discovery,lv2015observation}, NbAs\cite{xu2015discovery2}, and TaP \cite{xu2015observation} (and predicted in WTe$_2$, MoTe$_2$, and other Ta or Nb mono pnictides \cite {soluyanov2015new,sun2015prediction,weng2015weyl}). Very recently, Weyl nodes were shown in the intrinsically TRS-breaking compound, YbMnBi$_2$ \cite{borisenko2015time}. Young and Kane used the concept of non-symmorphic symmetry to predict that in 2D square lattices, new types of 2D DSMs can exist that are distinct from both graphene and 3D DSMs \cite{young2015dirac}. In particular, these 2D Dirac cones may host 2D Dirac fermions with a behavior distinct from their 3D analogues; experimental verification is pending. Finally, materials with Dirac line nodes, where the Fermi surface forms a closed loop, have recently been predicted but only experimentally verified in one material, PbTaSe$_2$, where other bands are interfering at the Fermi level \cite{PhysRevLett.115.036806,xie2015new,burkov2011topological,rhim2015landau,bian2015topological}.

\indent{} In all of the currently known DSMs, the energy range of the linear dispersion of the Dirac cone is very small. In Cd$_3$As$_2$, the Lifshitz transition appears according to calculations only about 20\,meV above the Fermi level. In the real material however, the Fermi energy has been shown to lie 200\,meV above the Dirac cone \cite{PhysRevLett.113.027603}. In Na$_3$Bi, TaAs and other monopnictides, the Lifshitz transition is only roughly 100\,meV away from crossings. A material with a larger linear dispersion would allow easy study of Dirac and Weyl physics despite changes in the Fermi level due to defects or impurities. The fabrication of thin films and devices from Dirac and Weyl materials would greatly benefit from more robust Dirac and Weyl states due to the difficulties in achieving thin film quality approaching that of single crystals. Also, many of the known materials have further disadvantages, such as the toxicity of arsenides as well as the extreme air sensitivity and chemical instability of Na$_3$Bi, which also make studying their exotic physics difficult.

\indent{} Here, we show by electronic structure calculations and ARPES that a so far unnoticed system, ZrSiS, exhibits several Dirac crossings within the Brillouin zone (BZ) which form a diamond shaped Fermi surface with a line of Dirac nodes, without any interference of other bands. This compound is non-toxic and highly stable with band dispersions of the linear region of those crossings being larger than in any other known compound: up to 2\,eV in some regions of the BZ. Spin orbit coupling (SOC) introduces a small gap to the Dirac cones near the Fermi surface, of the size of $\approx$ 20\,meV (much less than in the related Bi-based compounds). We also show the presence of a Dirac feature below the Fermi level, which is generated by the square Si sublattice and is protected by the non-symmorphic symmetry through a glide plane (regardless of SOC strength), supporting the recent prediction by Young and Kane regarding 2D Dirac fermions. We also show that around this Dirac feature an unusual, previously not predicted surface state arises. Thus, ZrSiS is a very promising candidate for investigating Dirac and Weyl physics, as well as the properties of lines of Dirac nodes.

\indent{} ZrSiS crystallizes in the PbFCl structure type in the tetragonal \textit{P4/nmm} space group (No. 129) \cite{klein1964zirconium}. It is related to the Weyl semimetal, YbMnBi$_2$, whose structure is a stuffed version of the PbFCl crystal structure, hence the different stoichiometry. Other Bi based, stuffed PbFCl structures, such as EuMnBi$_2$ and (Ca/Sr)MnBi$_2$, have also already been shown to host Dirac electrons and to exhibit exotic transport properties \cite{park2011anisotropic,may2014effect,lee2013anisotropic}. Both structures display square nets of Si and Bi atoms, respectively, that are located on a glide plane.



\begin{figure}[htb]
  \centering
  \includegraphics[width=13cm]{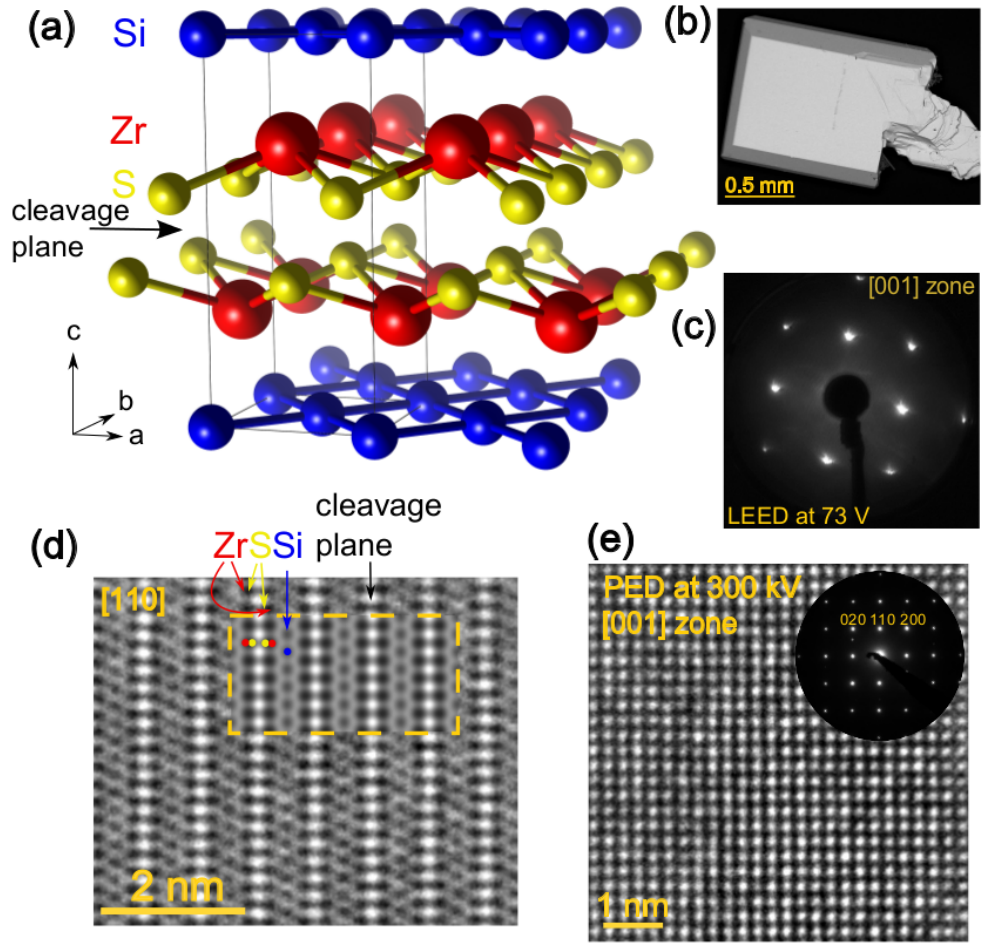}
  \caption{(a) Crystal structure of ZrSiS. The Si square net can be seen in blue. (b) SEM image of a typical crystal. (c) LEED pattern of a cleaved crystal showing that it cleaves perpendicular to the \textit{c} axis. (d) HRTEM image of the (110) surface, inset shows simulated HRTEM image. The focus plane is $\Delta$f = - 50 nm, close to the Scherzer focus, where atoms appear in black. Individual atoms could be identified and the cleavage plane between sulfur atoms is visible in white. For images with different foci and their simulations see SI. (e) HRTEM image and PED pattern of the (001) surface,  the square arrangement of atoms is clearly visible.}
\label{struc}
\end{figure}

\indent{} The crystal structure of ZrSiS is displayed in figure \ref{struc} (a). The Si square net is located in the \textit{ab} plane and layers of Zr and S are sandwiched between the Si square nets in such a way that there are neighboring S layers in between the Zr layers. We were able to image the square lattices with high resolution transmission electron microscopy (HRTEM) shown in figure \ref{struc} (d) and (e) (see SI for more information on precession electron diffraction (PED) and HRTEM). The HRTEM image of the (110) surface shows a gap between neighboring S atoms, which is where the crystals cleave. The LEED pattern shown in figure \ref{struc} (c) clearly indicates a square arrangement of Bragg reflections, hence showing that the crystals cleave perpendicular to the tetragonal \textit{c} axis. There is no sign of a surface reconstruction happening in these crystals. An SEM image of a typical crystal is shown in figure \ref{struc} (b). The crystals are very stable in water and air and only dissolve in concentrated acids.

\begin{figure}[htb]
  \centering
  \includegraphics[width=13cm]{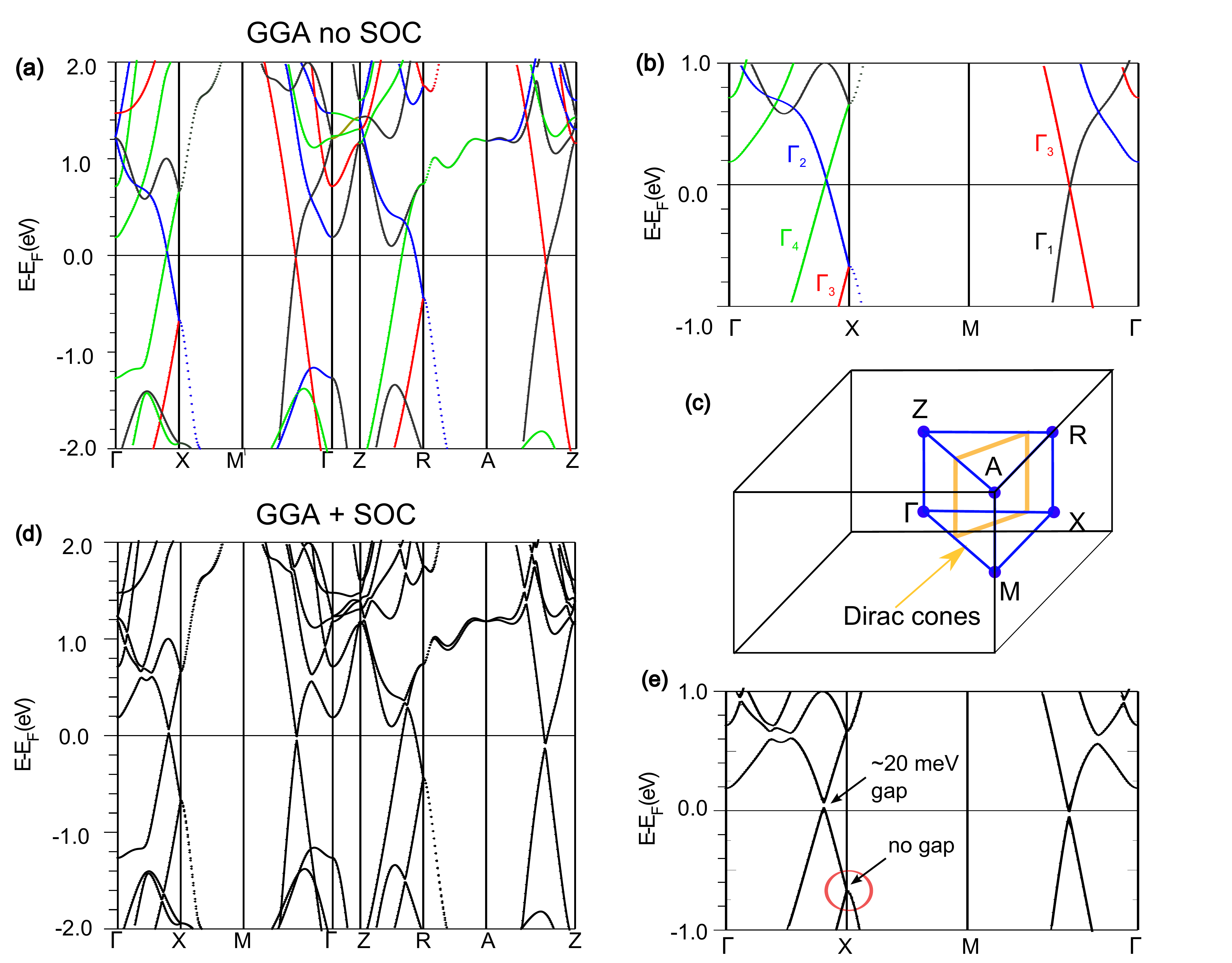}
  \caption{Calculated bulk band structure of ZrSiS. (a) Bulk band structure without SOC, each color of the line represents a different irreducible representation; they are labeled in the closed up image (b). (c) BZ of a primitive tetragonal lattice; the location of the Dirac cones along the symmetry lines is labeled in orange. (d) Bulk band structure with SOC; a very small gap is opened but the overall change to the band structure remains small. In the zoomed in panel (e) the cone at X, which is protected by non-symmorphic symmetry, is circled in red.}
\label{bulk}
\end{figure}

\indent{} The calculated electronic structure of bulk ZrSiS is displayed in figure \ref{bulk}. Without SOC, several Dirac cones are visible which cross along $\Gamma$X and $\Gamma$M as well as along ZR and ZA, which are the respective symmetry lines above $\Gamma$X and $\Gamma$M. However, the crossing along ZR is higher in energy compared to the one along $\Gamma$X. The Dirac cones form an unusually shaped line node in the BZ, displayed in figure \ref{bulk} (c) (this schematic assumes all Dirac points to be at the same energy to make it easier to understand the principle of the electronic structure). This gives rise to a diamond shaped Fermi surface. Note that the range in which the bands are linearly dispersed is very large compared to other known Dirac materials. The electronic structure without SOC is very similar to that observed in YbMnBi$_2$, however, since the symmetry along the lines with the Dirac crossings is $C_{2v}$, SOC gaps the cones. In Bi based compounds, this effect is very dramatic. It also destroys the large dispersion of the linear bands. In ZrSiS, however, SOC is small and only produces very small gaps in the cones along the $C_{2v}$ symmetry lines, maintaining the large linear dispersion (see figure \ref{bulk}). Furthermore, in Bi based compounds, more bands interfere with the cone structure around the Fermi energy.

 \indent{} There are other Dirac-like crossings at the X and R point that are located at $-0.7$\,eV and $-0.5$\,eV, respectively. These crossings are protected by the non-symmorphic symmetry of the space group, very similar to the recently predicted Dirac cones in 2D square nets and are not influenced by SOC \cite{young2015dirac}. In this template system, along the XM direction, both bands fold on the same energy. This degeneracy may subsequently be lifted to host 2D Dirac fermions. To the best of our knowledge this is the first time such a feature in the electronic structure has been observed in a real material. This Dirac-like crossing is significantly below the Fermi level with other bands also present, however, hole doping (on the Zr or S site) or the gating of thin films, may allow for investigation of the physics of the 2D Dirac fermions.

\begin{figure}[htb]
  \centering
 \includegraphics[width=13cm]{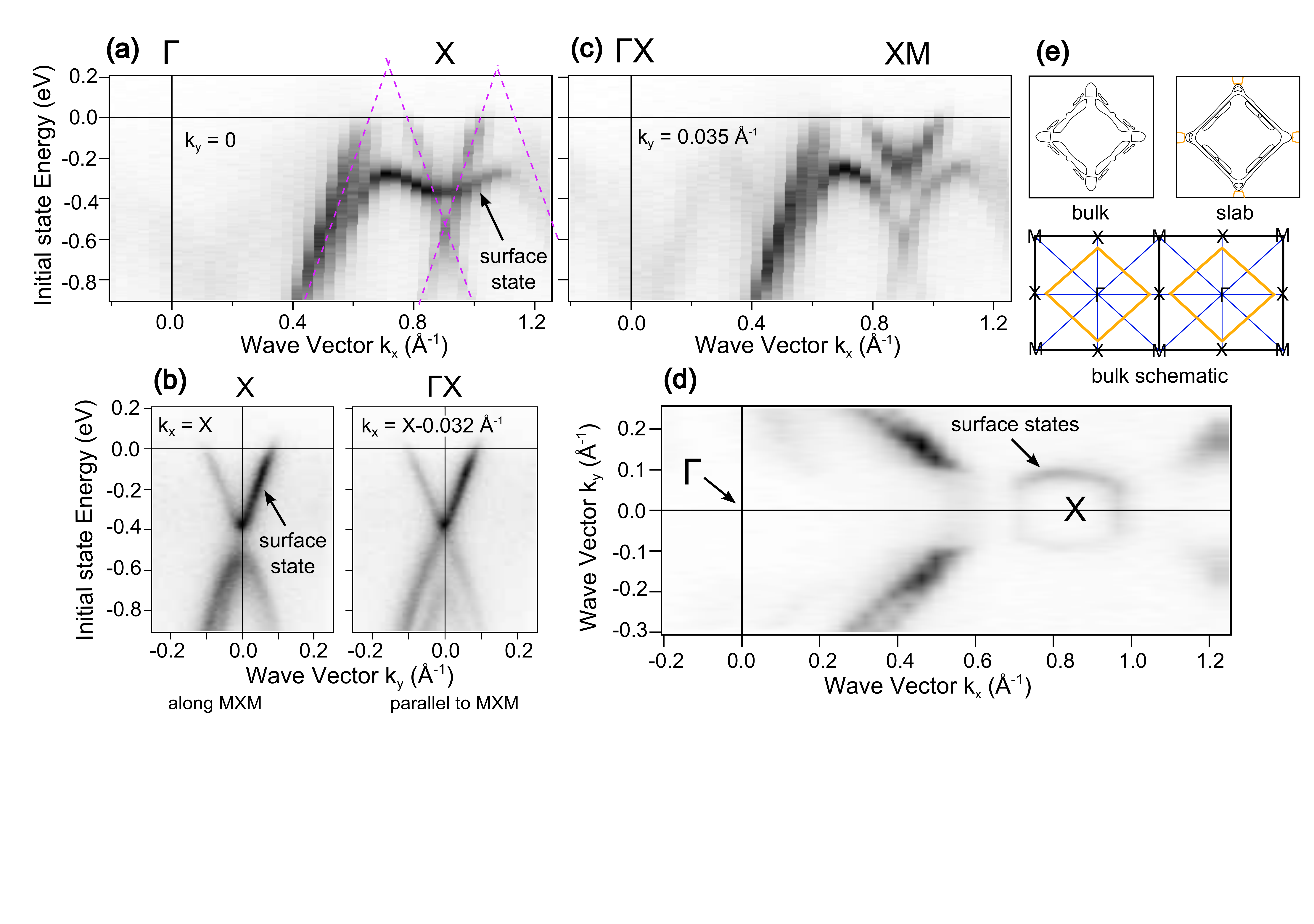}
  \caption{Band structure measured with ARPES. (a) Band structure along $\Gamma$X. The purple lines represent the predicted bulk bands. In addition a surface state is visible. (b) Band structure along MXM (left) and parallel to MXM (right). Along the high symmetry line the band structure is gapped (left panel) but with a much smaller gap than predicted in the bulk calculation. The gap closes and a cone forms parallel to the high symmetry line (right panel). (c) Band structure parallel to $\Gamma$X. Due to the gapping of the surface state it can be inferred that it is hybridized with the bulk cone at X.  (d) Constant energy plot at the Fermi energy. (e) The lower drawing sketches the predicted Fermi surface and compares calculated bulk and slab Fermi surfaces (upper panels). Pockets that are clearly surface derived are drawn in orange. The measured Fermi surface displays the predicted slab Fermi surface well.}
\label{arpes}
\end{figure}

\indent{} ARPES data are shown in figure \ref{arpes}.
The cone protected by non-symmorphic symmetry at X is clearly visible at $-0.5$\,eV (figure \ref{arpes} (a)). Perpendicular to $\Gamma$X, both bands fold on the same energy along the XM direction (left panel of figure \ref{arpes} (b)) exactly matching the prediction of Young and Kane \cite{young2015dirac}. The Dirac points of the cones along the $\Gamma$X line in figure \ref{arpes} (a) are not completely visible since the Fermi level is slightly below Dirac points (as also predicted in the slab calculation shown below). Dashed purple lines indicate the predicted bulk bands. Note that we observe linearly dispersed bands for more than 1\,eV energy range below the Fermi level, as predicted by the calculation.  We observe additional states along $\Gamma$X (figure \ref{arpes} (a)) not seen in the calculated bulk band structure, which we attribute to surface-derived states. When moving parallel to $\Gamma$X towards XM  (figure \ref{arpes} (c)), one can see how this surface state interacts with the bulk bands near X. 
This hybridization of the alleged surface state with the conical bulk state near the X point may be attributed to the inherent two-dimensional character of this bulk state. Normally, a surface state does not exist within the projected bulk band structure. However, as the bulk state is rather two-dimensional itself, we surmise that the bands hybridize in the vicinity of the surface. Since ARPES is an extremely surface sensitive technique and the unit cell along the c-axis is rather long, the actual bulk band dispersion inside the crystal remains unobservable. The left panel of figure \ref{arpes} (b) shows the measured band structure along MXM. Along the high symmetry line, a gap is observed, which is much smaller than the gap in the bulk band structure, due to the presence of surface states along this direction as well. Another cone at $-0.4$\,eV is visible in the measurement slightly parallel to the high symmetry line, towards the $\Gamma$X direction (right panel of figure \ref{arpes} (b)). This cone is not seen in the bulk band structure calculation (see figure S5), which indicates that this Dirac cone is also surface derived. It is connected to the surface state along $\Gamma$X as seen in figure \ref{arpes} (a) as well as in the slab calculation (see below). 
In figure \ref{arpes} (d), we show a constant energy plot of the Fermi level. From the bulk calculation, we expect a diamond shaped Fermi surface as sketched in figure \ref{arpes} (e) (lower panel). In the experimental data, we not only observe this diamond shaped Fermi surface, but, in addition, the data shows the surface-derived state around X to cross the Fermi level as well. The calculated slab Fermi surface in figure \ref{arpes} (e) (upper panel) is in excellent agreement with the measured Fermi surface. If a constant energy plot is taken at lower energies (see figure S3 in the SI) the observed ARPES spectrum matches very well with the predicted constant energy surface at -515\,meV of the bulk surface, due to the absence of surface states at this energy.

\begin{figure}[htb]
  \centering
  \includegraphics[width=13cm]{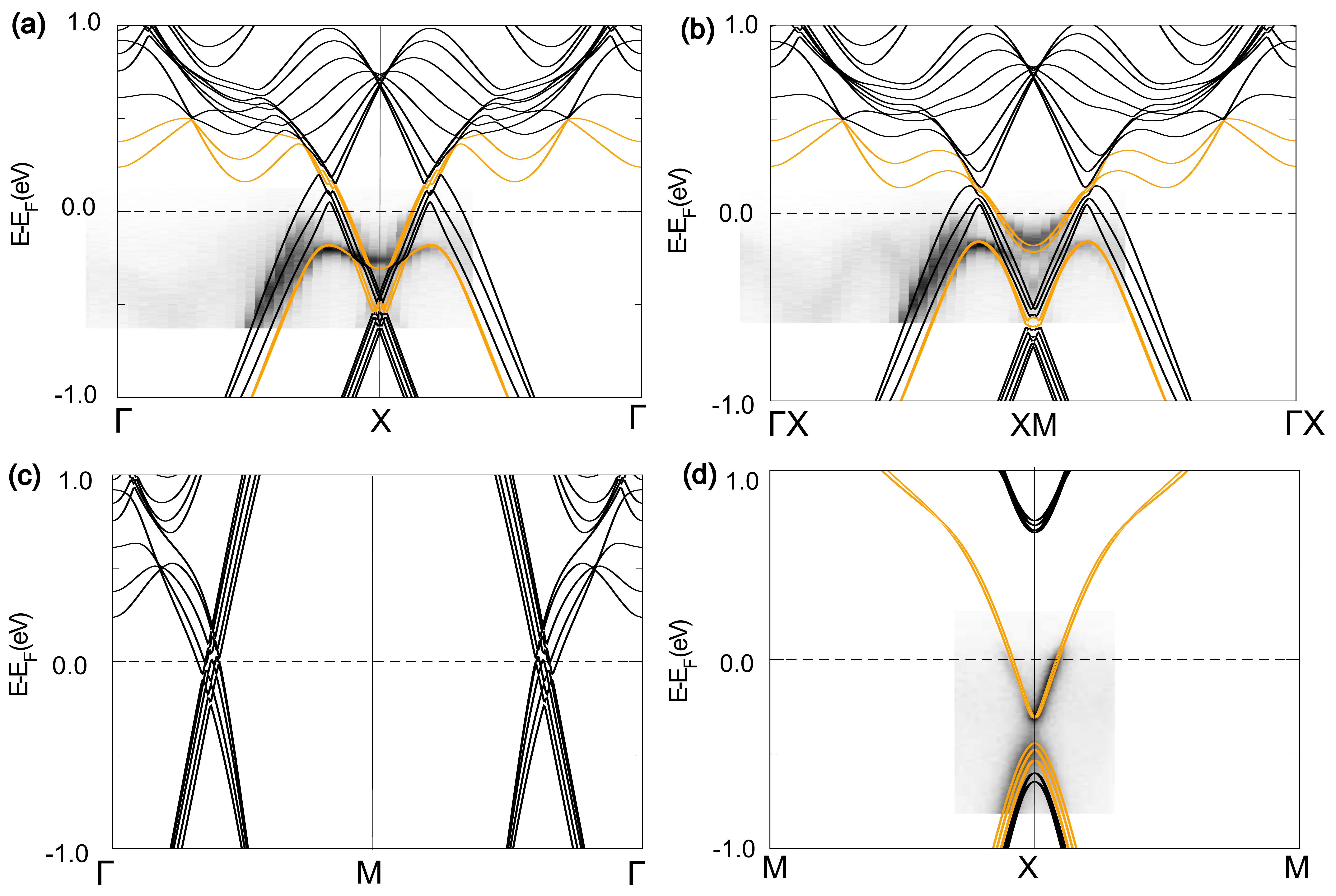}
  \caption{Calculated slab band structure in comparison with the measured band structure. (a) Bands along $\Gamma$X. The orange bands indicate how bands are progressing, showing the hybridization of the surface state with the bulk. (b) Bands parallel to $\Gamma$X highlighting the mixing of surface states and bulk bands belonging to the cone at X. (c) Slab band structure along $\Gamma$M, showing that there is no big change to the bulk band structure in this direction. (d) Bands along MXM, surface bands are highlighted in orange. The surface states significantly reduce the bulk gap along this direction.}
\label{slab}
\end{figure}

\indent{} In order to ascertain the nature of the alleged surface states, we performed band structure calculations of a slab. The resulting band structures in comparison to ARPES data are shown in figure \ref{slab}. SOC is included for these calculations. The creation of a surface causes several changes to the electronic structure; the cone along $\Gamma$M remains unchanged (figure \ref{slab} (c)) but the cone along $\Gamma$X moves up in energy compared to the bulk structure. The same is true for the cone at X (see figure \ref{slab} (a)). This can be understood if one considers that in a 2D slab, the ZR bands are projected onto the $\Gamma$X bands. In addition, the surface state seen in ARPES appears along $\Gamma$X. The high level of agreement between the predicted and measured electronic structure is shown by superimposing the two images in the figure without rescaling. Furthermore, the measured Fermi energy matches the predicted one, however, the ARPES data are measured at room temperature which indicates that the samples are slightly hole doped. 

\indent{} Continuous bands along this surface state are highlighted in orange in figure \ref{slab} (a). These bands do not follow the expected path of the surface state, thus also indicating the hybridization with the bulk. A simulation of the slab band structure parallel to $\Gamma$X is shown in figure \ref{slab} (b). In accordance with ARPES the bands that form the surface state split apart.
 Calculations of the slab without SOC (see figure S4 (a) in the SI) show that the bands forming the surface state and the cone around X have different irreducible representations in the absence of SOC, along $\Gamma$X, but not parallel to it (figure S4 (b)). This indicates that SOC cannot be held responsible for the hybridization. Figure S4 (c) in the SI shows the contribution of the surface atoms to the slab band structure. 
This supports the surface character of the additional band observed in the experiment. 

\indent{} Figure \ref{slab} (d) shows the predicted slab band structure along XM. Surface states that lie in between the bulk band gap in this part of the BZ are highlighted in orange. The surface states are mainly appearing around the X point. Again, ARPES data matches well with the prediction, the observed decreasing gap along this direction is also seen in the slab calculation showing that this is caused by the surface.


\indent{} In summary, we showed that ZrSiS, a stable and non toxic material, has a very exotic electronic structure with many Dirac cones that form a diamond shaped Fermi surface. The bands are linearly dispersed over a very large energy range, larger than in any other material reported to date. We confirmed our electronic structure calculations with ARPES measurements that are in excellent agreement with the calculated structure. We also show the first experimental realization of template system for 2D Dirac cones protected by non-symmorphic symmetry, in excellent agreement with recent theoretical predictions. In addition, we observe an unconventional surface state that is hybridized with bulk bands around the X point. It is uncommon for a surface state to exist within the projected bulk band structure or even hybridize with it. A possible cause might be the 2D nature of the bulk bands around X, however, further investigation into this effect is required. In contrast to compounds with a Bi square net, where a large SOC opens a large gap with parabolic dispersion, ZrSiS has a Si square net where the SOC effect is very much reduced and linear dispersion of the bands is mostly preserved. Since no other bands interfere at the Fermi level, the unusual electronic structure of ZrSiS makes it a strong candidate for further studies into Dirac and Weyl physics; especially magnetotransport, since the Fermi energy can be tuned quite substantially, while still being in the linear range of the bands.

\section{methods}

\indent{} Single crystals of ZrSiS were grown in a two step synthesis. First, a polycrystalline powder was obtained following the same procedure as in \cite{klein1964zirconium}. In a second step single crystals were grown from the polycrystalline powder via I$_2$ vapor transport at 1100$^\circ$C with a 100$^\circ$C temperature gradient. The crystals were obtained at the cold end. The published crystal structure was confirmed with single crystal x-ray diffraction and electron diffraction. The crystal used for SXRD was of extremely high quality and an R$_1$ value of 1.5\% was obtained for the structural solution (see table S1 and S2 in the supplemental information (SI) for more details). Single crystal x-ray diffraction data was collected  on a STOE IPDS II working with graphite monochromated Mo K$_\alpha$ radiation. Reflections were integrated with the STOE X -Areaa 1.56 software and the structure was solved and refined by least square fitting using SHELXTL \cite{Sheldrick:sc5010}. Electron microscopy was performed with a Phillips CM30 ST (300 kV, LaB$_6$ cathode). High resolution transmission microscopy (HRTEM) images and precession electron diffraction (PED) patterns were recorded with a CMOS camera (TemCam-F216, TVIPS) equipped with a nanomegas spinning star to obtain PED images. The program JEMS (Staddmann) was used to simulate diffraction patterns and HRTEM micrographs. For ARPES measurements crystals were cleaved and measured in ultra high vacuum (low 10$^{-10}$ mbar range). Low energy electron diffraction (LEED) showed that the cleavage plane was the (001) plane. ARPES spectra where recorded at room temperature with a hemispherical PHOIBOS 150 electron analyzer (energy and angular resolution are 15 meV and 0.5$^\circ$, respectively). As photon source a monochromatized He lamp that offers UV radiation at $h\nu$ = 21.2 eV (He I) was used. SEM images of crystals were measured with a scanning electron microscope (SEM; Vega TS 5130 MM, Tescan) using a Si/Li detector (Oxford). 

\indent{}Electronic structure calculations were performed in the framework of density functional theory (DFT) using the \textsc{wien2k} \cite{blaha2001} code with a full-potential linearized augmented plane-wave and local orbitals [FP-LAPW + lo] basis \cite{singh2006} together with the Perdew-Becke-Ernzerhof (PBE) parameterization \cite{perdew_generalized_1996} of the Generalized Gradient Approximation (GGA) as the exchange-correlation functional. The plane wave cut-off parameter R$_{MT}$K$_{MAX}$ was set to 7 and the irreducible Brillouin zone was sampled by 1368 k-points (bulk) and by a 30x30x3 mesh of k-points (slab). Experimental lattice parameters from the single crystal diffraction studies were used in the calculations. Spin orbit coupling (SOC) was included as a second variational procedure. For the slab calculation it was found that cleaving between sulfur atoms resulted in the closest match to the experimental observation. This cleavage plane is in agreement with HRTEM imaging and chemical intuition. The slab was constructed by stacking 5 unit cells in \textit{c} direction that were gapped by a 5.3 \AA~ vacuum.

\clearpage

\bigskip 
\begin{acknowledgments}
The authors would like to thank Roland Eger for single crystal diffraction measurements. LMS gratefully acknowledges funding from the Minerva fast track fellowship by the Max Planck society. 
\end{acknowledgments}

\section{Author Contribution}
LMS initiated the project, grew the crystals and calculated the electronic structures. LMS, MNA, and CRA analyzed the results, interpreted them, and put them into context. CS measured ARPES spectra. VD performed HRTEM and PED analysis. SSP, BVL, and CRA supervised the research. All authors discussed the results and contributed to writing the manuscript.


\clearpage

\clearpage

\clearpage

\clearpage

\clearpage

\end{document}



\title{Supplementary Information for "Dirac Cone Protected by Non-Symmorphic Symmetry and 3D Dirac Line Node in ZrSiS"}

\author{Leslie M. Schoop}
\email{l.schoop@fkf.mpg.de}
\affiliation{Max Planck Institute for Solid State Research, Heisenbergstr. 1, 70569 Stuttgart, Germany}
\author{Mazhar N. Ali}
\affiliation{Max Plank Institute for Microstructure Physics, Weinberg 2, 06120 Halle, Germany}
\affiliation{IBM-Almaden Research Center, 650 Harry Road, San Jose, CA 95120, USA}
\author{Carola Stra\ss er}
\affiliation{Max Planck Institute for Solid State Research, Heisenbergstr. 1, 70569 Stuttgart, Germany}
\author{Viola Duppel}
\affiliation{Max Planck Institute for Solid State Research, Heisenbergstr. 1, 70569 Stuttgart, Germany}
\author{Stuart S. P. Parkin}
\affiliation{Max Plank Institute for Microstructure Physics, Weinberg 2, 06120 Halle, Germany}
\affiliation{IBM-Almaden Research Center, 650 Harry Road, San Jose, CA 95120, USA}
\author{Bettina V. Lotsch}
\email{b.lotsch@fkf.mpg.de}
\affiliation{Max Planck Institute for Solid State Research, Heisenbergstr. 1, 70569 Stuttgart, Germany}
\affiliation{Department of Chemistry, Ludwig-Maximilians-Universit\"at M\"unchen, Butenandtstr. 5-13, 81377 M\"unchen, Germany}
\affiliation{Nanosystems Initiative Munich (NIM) \& Center for Nanoscience, Schellingstr. 4, 80799 M\"unchen, Germany}
\author{Christian R. Ast}
\affiliation{Max Planck Institute for Solid State Research, Heisenbergstr. 1, 70569 Stuttgart, Germany}
\date{\today}


\pacs{}

\maketitle

\indent{} 
 
 \begin{table}[htbp]
\caption{Crystallographic data and details of data collection.}
\begin{center}
\begin{tabular}{cc}
\hline
\hline
Formula Sum & ZrSiS \\[1ex]
\hline
\hline 
Crystal system  & Tetragonal \\
space group & \textit{P4/nmm} (No. 129) \\ 
Formula weight (g/mol) & 151.37  \\
Density (g/cm$^3$) & 4.964  \\
\textit{a} (\AA) & 3.5450(5)  \\ 
\textit{c} (\AA) & 8.0578(16)  \\ 
\textit{V} (\AA$^3$) & 101.27(3)   \\ 
\textit{Z} &  2  \\ 
Temperature (K) & 293(2)  \\ 
Radiation & Mo K$_\alpha$ \\
\textit{2 $\Theta$} range & 10.13 - 63.36 \\
F$_{000}$ &  140 \\
Unique reflections & 132 \\
Difference e$^-$ density (e/\AA$^3$) & +0.917 to -1.061 \\
Extinction coefficient &  0.087(11) \\
R$_1$ (all reflections) & 0.0205  \\
R$_1$ (F$_0$ $>$ 4$\sigma$(F$_0$)) & 0.0153 \\
wR$_2$ & 0.0375 \\
Absorption coefficient $\mu$ (mm$^{-1})$ & 3.27 \\
GooF & 1.18 \\
\hline
\hline
\end{tabular}
\end{center}
\label{tab1}
\end{table}
 
  \begin{table}[htbp]
\caption{Position coordinates and thermal displacement parameters for ZrSiS}
\begin{center}
\begin{tabular}{ccccccccc}
\hline
\hline 
Atom & Wyck & x & y & z & Occ. & U$_{11}$ [\AA{}$^2$] & U$_{22}$ [\AA{}$^2$]  & U$_{33}$ [\AA{}$^2$]\\[1ex]
Zr & 2c & 0.25000 & 0.25000 & 0.72906(6)   & 1 & 0.0048(2) & 0.0048(2) & 0.0052(3) \\ 
S & 2c & 0.25000 & 0.25000 & 0.37915(15)   & 1 & 0.0057(4) & 0.0057(4) & 0.00715(5) \\
Si & 2a & 0.25000 & -0.25000 & 0.0000   & 1 & 0.0060(4)  & 0.0060(4) & 0.0062(6)\\
\hline
\hline
\end{tabular}
\end{center}
\label{tab2}
\end{table}

 \indent{} 

\begin{figure}[htb]
  \centering
  \includegraphics[width=13cm]{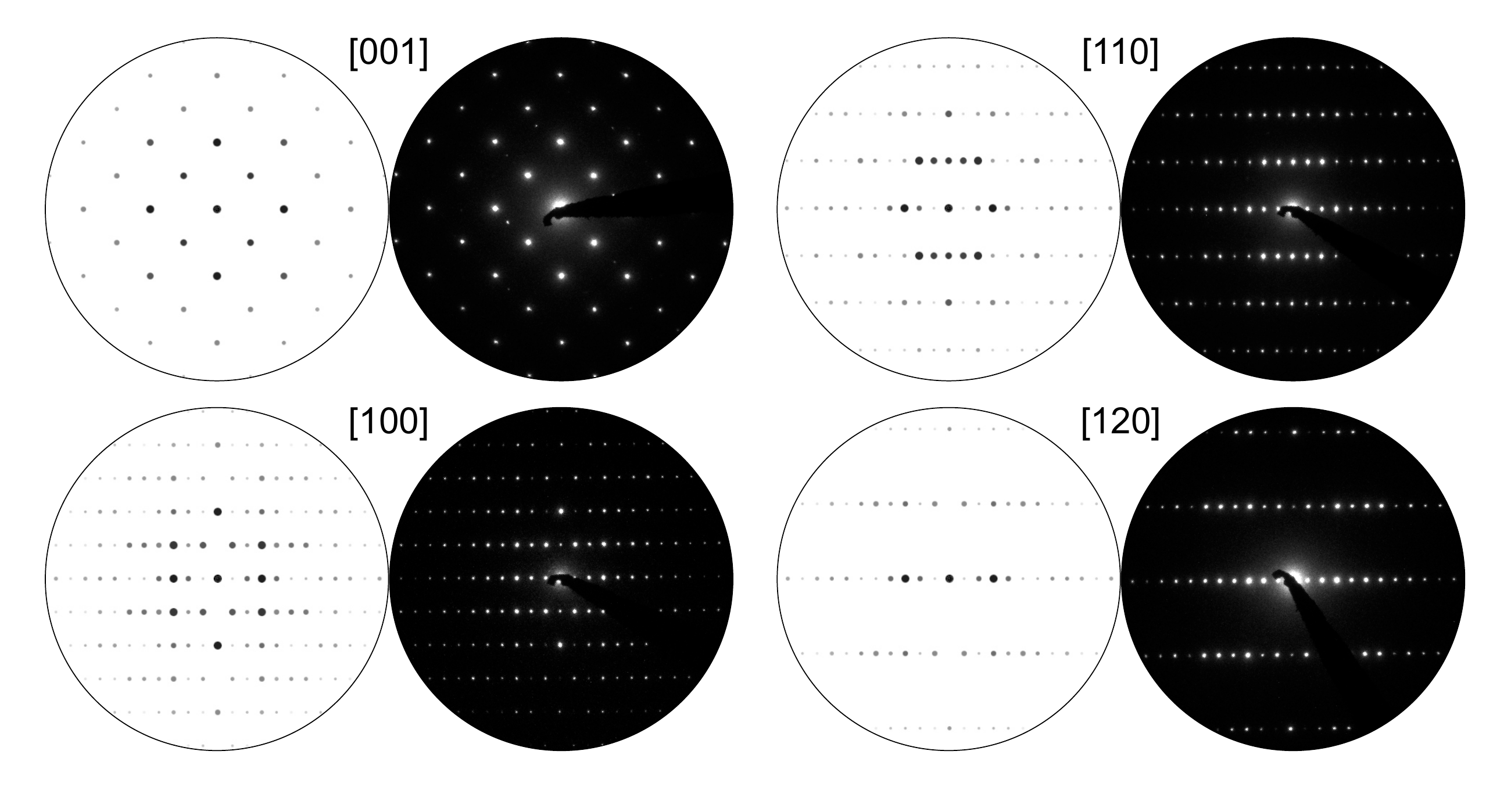}
  \caption{PED of ZrSiS, simulated patterns are shown on the left and measured patterns on the right.}
\label{PED}
\end{figure}
\clearpage
\begin{figure}[htb]
  \centering
  \includegraphics[width=13cm]{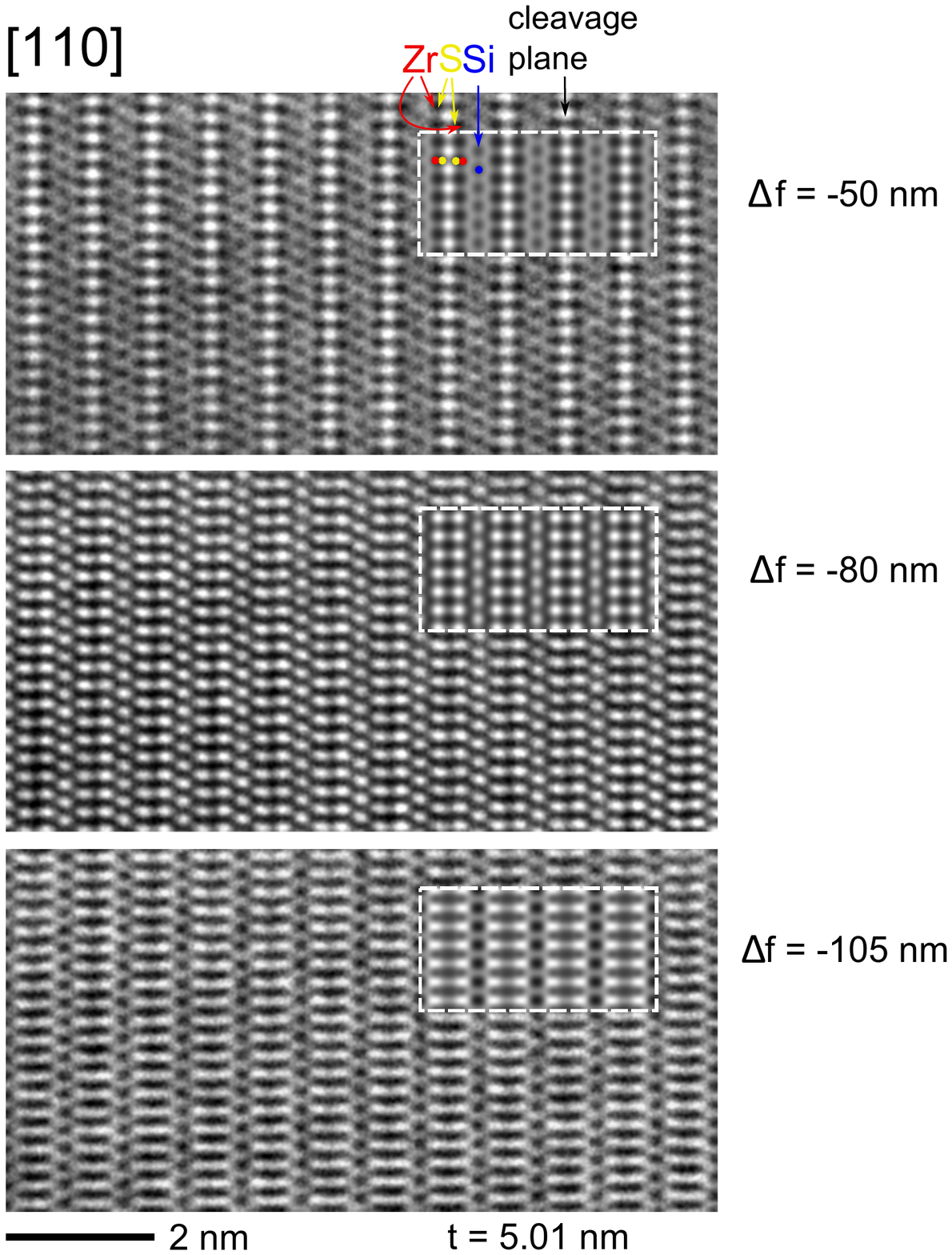}
  \caption{HRTEM images of the (110) plane of ZrSiS, measured on a powdered sample. Top panel is close to Sherzer focus, where atoms appear in black and gaps in white. Individual atoms are labeled. Insets show simulated HRTEM images; simulations assume a sample thickness of t = 5.01nm.}
\label{hrem-110}
\end{figure}
\clearpage
\begin{figure}[htb]
  \centering
  \includegraphics[width=13cm]{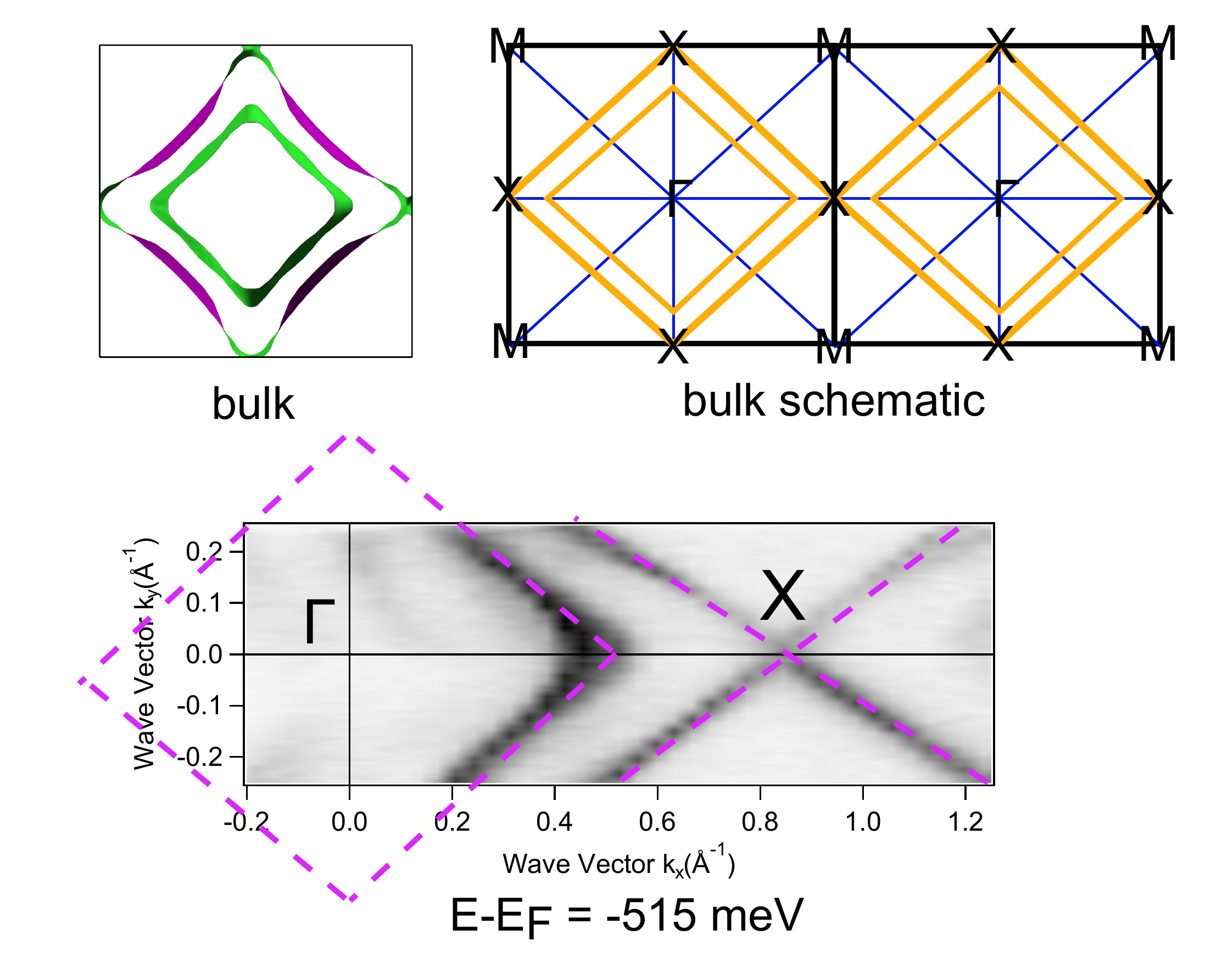}
  \caption{Constant energy surface at E-E$_F$ = -515 meV. The energy surface now expands to the X point as expected. A second diamond shaped surface appears inside, which is the second branch of the bulk cones. The drawing on top sketches the predicted energy surface at -515 meV from bulk calculations (right), whereas the left shows the actually calculated bulk energy surface. Dashed purple lines indicate the measured bulk bands. Note that the crystal was not in a perfect 90 degree angle to the beam which is why the measured bands do not perfectly map the diamond shaped energy surface.}
\label{EF-515ev}
\end{figure}
\clearpage

\begin{figure}[htb]
  \centering
  \includegraphics[width=10cm]{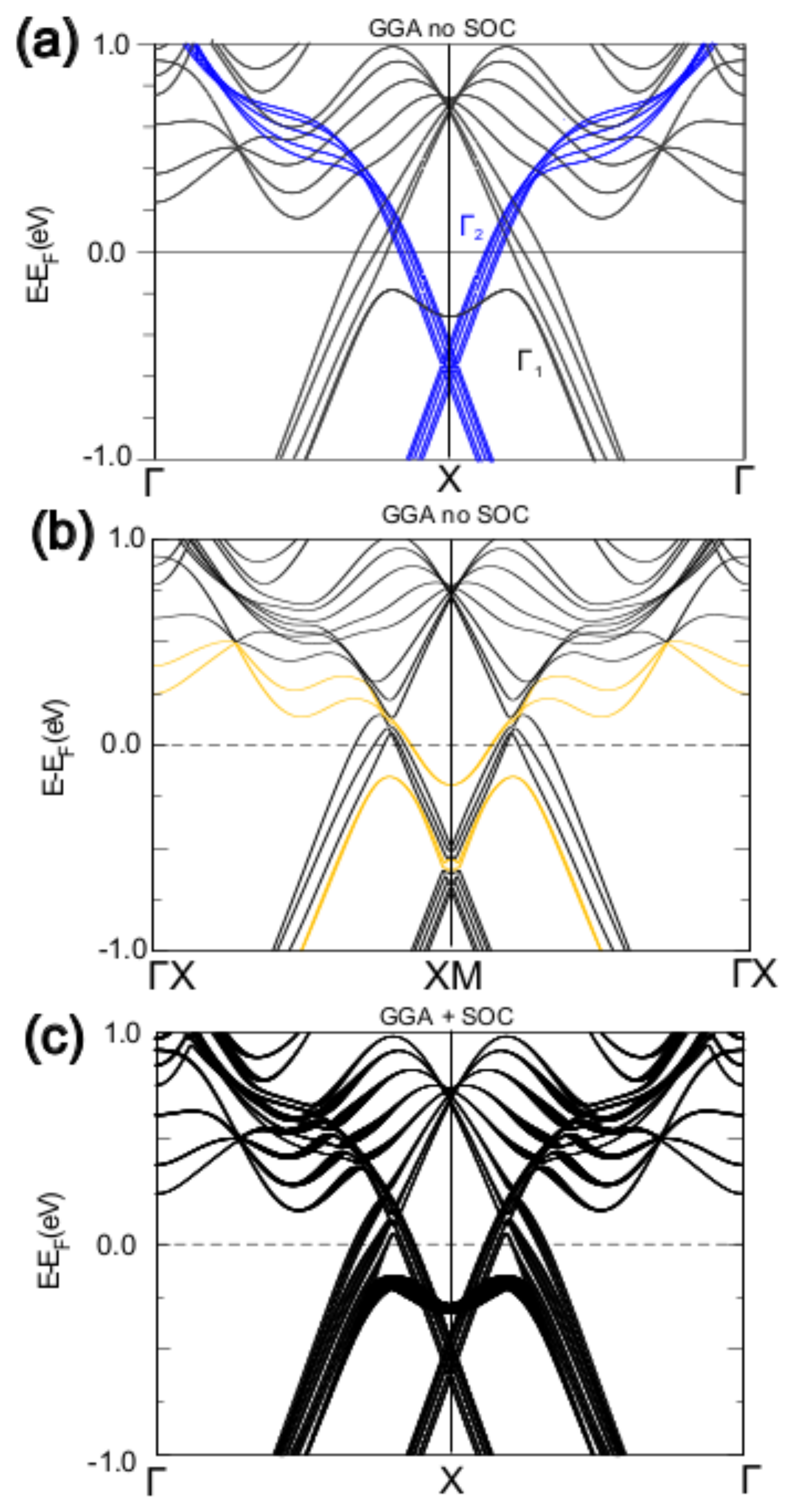}
  \caption{Slab band structure along $\Gamma$X (a) calculated without spin orbit coupling showing that two different irreducible representations are possible and the surface state does not mix with the bulk bands. (b) Calculated without SOC and plotted parallel to $\Gamma$X where only one irreducible representation is allowed and hence mixing of the states is allowed, even without SOC. (c) Calculated with SOC but contribution of surface atoms is plotted thick, indicating clearly that the surface state arises from these atoms even if it is hybridized with the bulk.}
\label{slab_fatbands}
\end{figure}
\clearpage
\begin{figure}[htb]
  \centering
  \includegraphics[width=13cm]{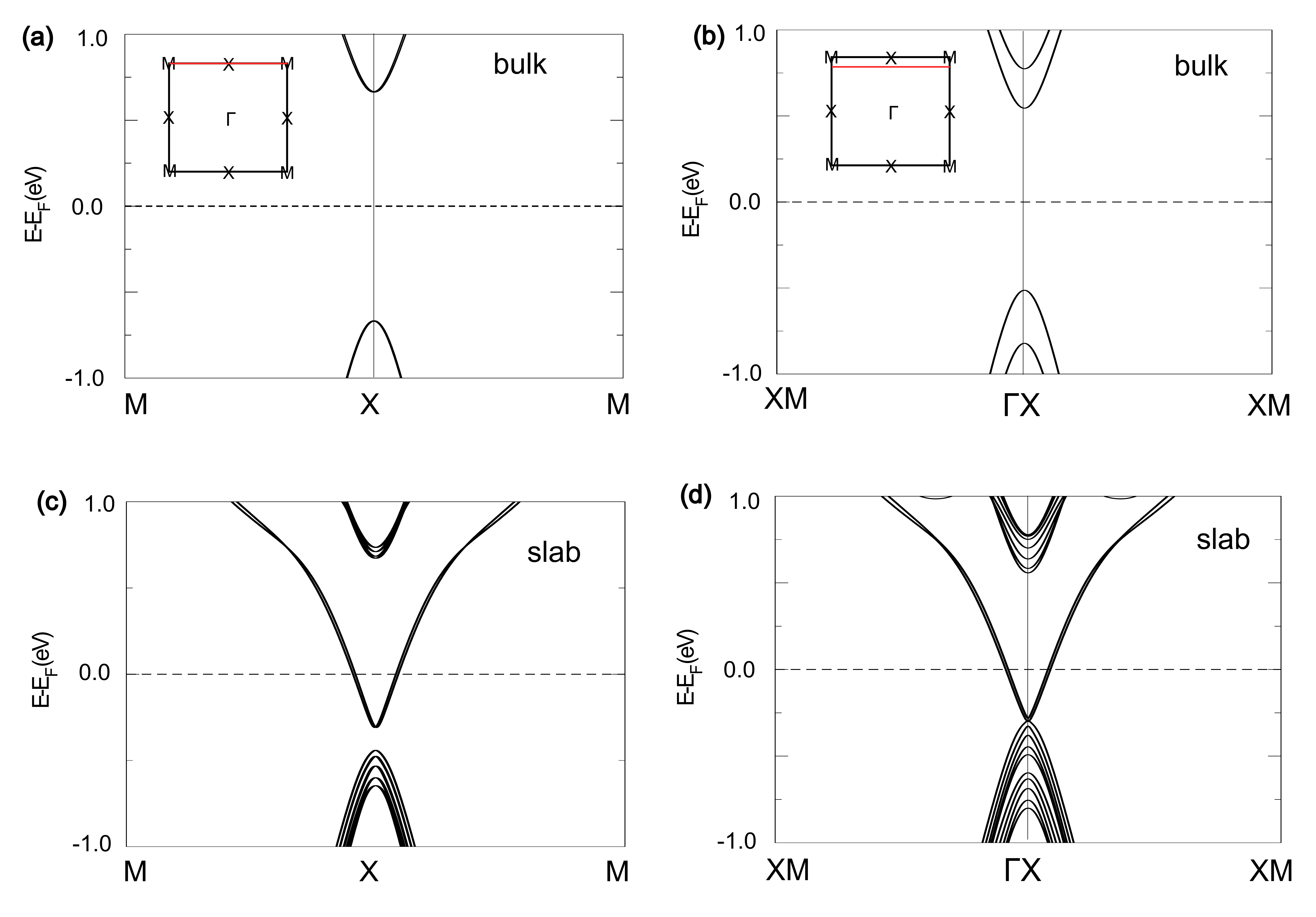}
  \caption{Band structure along MX and parallel to MX. (a) and (b) show bulk band structures where the BZ is gapped along these directions, insets show the k path. (c) and (d) show slab band structures along the same paths, the gap is closed by surface states giving rise to a cone that is also observed in ARPES.}
\label{MXM}
\end{figure}
\clearpage